\documentclass{aa}  

\usepackage{graphicx}
\usepackage{txfonts}
\begin{document}

   \title{Establishing the accuracy of asteroseismic mass and radius estimates of giant stars\thanks{Based on observations made with the Nordic Optical Telescope, owned in collaboration by the University of Turku and Aarhus University, and operated jointly by Aarhus University, the University of Turku, and the University of Oslo, representing Denmark, Finland, and Norway, the University of Iceland, and Stockholm University at the Observatorio del Roque de los Muchachos, La Palma, Spain, of the Instituto de Astrofisica de Canarias.}}

   \subtitle{III. KIC4054905, an eclipsing binary with two 10 Gyr thick disk RGB stars}

   \author{K. Brogaard
          \inst{1, 2}
          \and
          T. Arentoft\inst{1}
          \and
          D. Slumstrup\inst{3,1}
          \and
          F. Grundahl\inst{1,2}
          \and
          M. N. Lund\inst{1,2}
          \and
          L. Arndt \inst{4} 
         \and 
         S. Grund\inst{4}
          \and
          J. Rudrasingam\inst{4}
          \and
          A. Theil\inst{4}
          \and
          K. Christensen\inst{4}
          \and
          M. Sejersen\inst{4}
         \and
          F. Vorgod \inst{4}
          \and
          L. Salmonsen \inst{4}
         \and 
         L. \O rtoft Endelt \inst{4}
         \and
         S. Dainese \inst{4}
          \and
         S. Frandsen\inst{1}
         \and
         A.\,Miglio\inst{5,6,7,1}
         \and
         J. Tayar\inst{8}
         \and
         D. Huber\inst{9}
          }

   \institute{Stellar Astrophysics Centre, Department of Physics and Astronomy, Aarhus University, Ny Munkegade 120, DK-8000 Aarhus C, Denmark
\and
Astronomical Observatory, Institute of Theoretical Physics and Astronomy, Vilnius University, Saul\.{e}tekio av. 3, 10257 Vilnius, Lithuania
\and
European Southern Observatory, Alonso de Cordova 3107, Vitacura, Santiago, Chile
\and
Department of Physics and Astronomy, Aarhus University, Ny Munkegade 120, 8000 Aarhus C, Denmark
\and
Dipartimento di Fisica e Astronomia, Universit\`{a} degli Studi di Bologna, Via Gobetti 93/2, I-40129 Bologna, Italy \and
INAF – Osservatorio di Astrofisica e Scienza dello Spazio di Bologna, Via Gobetti 93/3, I-40129 Bologna, Italy
\and
School of Physics and Astronomy, University of Birmingham, Edgbaston, B15 2TT, UK
\and
Department of Astronomy, University of Florida, Bryant Space Science Center, Stadium Road, Gainesville, FL 32611, USA
\and
Institute for Astronomy, University of Hawai`i, 2680 Woodlawn Drive, Honolulu, HI 96822, USA
}

   \date{Received September XX, XXXX; accepted March YY, YYYY}

 
  \abstract
   {Eclipsing binary stars with an oscillating giant component allow accurate stellar parameters to be derived and asteroseismic methods to be tested and calibrated. To this aim, suitable systems need to be firstly identified and secondly measured precisely and accurately. KIC\,4054905 is one such system, which has been identified, but with measurements of a relatively low precision and with some confusion regarding its parameters and evolutionary state.}
   {Our aim is to provide a detailed and precise characterisation of the system and to test asteroseismic scaling relations.}
   {Dynamical and asteroseismic parameters of KIC4054905 were determined from \textit{Kepler} time-series photometry and multi-epoch high-resolution spectra from FIES at the Nordic Optical Telescope.}
   {KIC\,4054905 was found to belong to the thick disk and consist of two lower red giant branch (RGB) components with nearly identical masses of $0.95 M_{\odot}$ and an age of $9.9\pm0.6$ Gyr. The most evolved star with $R\simeq8.4 R_{\odot}$ displays solar-like oscillations. These oscillations suggest that the star belongs to the RGB, supported also by the radius, which is significantly smaller than the red clump phase for this mass and metallicity. Masses and radii from corrected asteroseismic scaling relations can be brought into full agreement with the dynamical values if the RGB phase is assumed, but a best scaling method could not be identified.}
   {The dynamical masses and radii were measured with a precision better than 1.0\%.
   We firmly establish the evolutionary nature of the system to be that of two early RGB stars with an age close to 10 Gyr, unlike previous findings. The metallicity and Galactic velocity suggest that the system belongs to the thick disk of the Milky Way.
    
   We investigate the agreement between dynamical and asteroseismic parameters for KIC\,4054905 measured in various ways. This suggests that consistent solutions exist, but the need to analyse more of these systems continues in order to establish the accuracy of asteroseismic methods.}


   \maketitle
%

\section{Introduction}

Asteroseismology offers the potential for new insights into stars, planets, and our Galaxy through the exploitation of high-precision photometric time series from space missions such as {\it Kepler} \citep{Borucki2010}, K2 \citep{Howell2014}, and TESS \citep{Ricker2014}. However, in order to ensure a correct interpretation of the rapidly increasing amounts of observational data, we still need to establish the accuracy level of the asteroseismic methods.

Many studies rely on the so-called asteroseismic scaling relations involving the frequency of maximum power, $\nu_{\mathrm{max}}$, and the large frequency spacing between modes of the same degree, $\Delta \nu$. On the one hand, $\Delta \nu$ has been shown to scale approximately with the mean density of a star \citep{Ulrich1986}; on the other hand, $\nu_{\mathrm{max}}$ scales approximately with the acoustic cut-off frequency of the atmosphere, which is related to surface gravity and effective temperature \citep{Brown1991,Kjeldsen1995,Belkacem2011}. In equation form, these relations are the following:

\begin{eqnarray}\label{eq:01}
\frac{\Delta \nu}{\Delta \nu _{\odot}} & = & f_{\Delta \nu}\left(\frac{\rho}{\rho_{\odot}}\right)^{1/2},\\
\label{eq:02}
\frac{\nu _{\mathrm{max}}}{\nu _{\mathrm{max,}\odot}} & = & f_{\nu _{\mathrm{max}}} \frac{g}{g_{\odot}}\left(\frac{T_{\mathrm{eff}}}{T_{\mathrm{eff,}\odot}}\right)^{-1/2}.
\end{eqnarray}

Here, $\rho$, $g$, and $T_{\rm eff}$ are the mean density, surface gravity, and effective temperature, and we have adopted the notation of \citet{Brogaard2018} that includes the correction functions $f_{\Delta \nu}$ and $f_{\nu _{\mathrm{max}}}$. We have, however, changed the approximate equal sign to an equal sign, which is the correct way of presenting the equations assuming that the correction functions are accurate. By rearranging the equations, expressions for the mass and radius can be obtained instead: 

\begin{eqnarray}\label{eq:03}
\frac{M}{\mathrm{M}_\odot} & = & \left(\frac{\nu _{\mathrm{max}}}{f_{\nu _{\mathrm{max}}}\nu _{\mathrm{max,}\odot}}\right)^3 \left(\frac{\Delta \nu}{f_{\Delta \nu}\Delta \nu _{\odot}}\right)^{-4} \left(\frac{T_{\mathrm{eff}}}{T_{\mathrm{eff,}\odot}}\right)^{3/2},\\
\label{eq:04}
\frac{R}{\mathrm{R}_\odot} & = & \left(\frac{\nu _{\mathrm{max}}}{f_{\nu _{\mathrm{max}}}\nu _{\mathrm{max,}\odot}}\right) \left(\frac{\Delta \nu}{f_{\Delta \nu}\Delta \nu _{\odot}}\right)^{-2} \left(\frac{T_{\mathrm{eff}}}{T_{\mathrm{eff,}\odot}}\right)^{1/2}. 
\end{eqnarray}

Quite a few empirical tests of these and similar equations have been performed \citep{Brogaard2012, Miglio2012, Handberg2017, Huber2017, Brogaard2018, Benbakoura2021}, but a much more substantial effort is needed to establish the obtainable accuracy in general. Precise and accurate observations spanning a range in stellar parameters are needed because $f_{\Delta \nu}$, and potentially also $f_{\nu _{\mathrm{max}}}$, are non-linear functions of the stellar parameters. The solar reference values, which we adopt in this work to be $\Delta \nu _{\odot} = 134.9\,\mu$Hz and $\nu _{\mathrm{max,}\odot} = 3090\,\mu$Hz following \citet{Handberg2017}, are also subject to uncertainties, which further complicates tests of the correction factors. We adopted an effective temperature of the Sun, $T_{\mathrm{eff,}\odot}=5772$\,K, according to IAU
2015 Resolution B3 \citep{IAU2015}; however, even this number varies among investigations, though at a very low level. 

Some works attempt to instead calibrate the asteroseismic scaling relations using the observations. However, perhaps due to the small number of measured systems, many of these calibrations ignore the known dependence of the correction functions on the effective temperature, metallicity, and/or evolutionary state by assuming $f_{\Delta \nu}=1$ and only calibrate the solar reference values or a constant offset, as can be seen \citet{Mosser2013}, \citet{Themessl2018}, and \citet{Benbakoura2021}, for example. 

The known dependency of $f_{\Delta \nu}$ on the stellar temperature, metallicity, and mass was first demonstrated using models by \citet{White2011} and later, in more detail and including core-helium-burning stars, by \citet{Miglio2013}, \citet{Sharma2016}, and \citet{Rodrigues2017}, for example. \citet{Guggenberger2017} also published predictions for $f_{\Delta \nu}$, but not for the core-helium-burning phase. \citet{Handberg2017} showed that the trends for $f_{\Delta \nu}$ seen in models are in agreement with observations. Therefore, calibrations that ignore this cannot be accurate.

Unlike $\Delta \nu$, $\nu _{\mathrm{max}}$ is not yet understood to a level where it can be modelled directly; although, some efforts have been made to obtain a physical understanding of it \citep{Belkacem2011}. \citet{Viani2017} suggest that $f_{\nu _{\mathrm{max}}}$ might include the stellar mean molecular weight, $\mu$, and the adiabatic exponent, $\Gamma_1$, so that 
\begin{equation}
f_{\nu _{\mathrm{max}}}\simeq\left(\frac{\mu}{\mu_\odot}\right)^{1/2}\left(\frac{\Gamma_1}{\Gamma_{1,\odot}}\right)^{1/2},
\end{equation}
but no empirical tests have been conducted yet.

\citet{Li2022} attempted to carry out a calibration of the scaling relations based on direct asteroseismic frequency modelling, suggesting a metallicity dependence in the relation between $\nu _{\mathrm{max}}$ and the surface gravity, which could relate to the theoretical work of \citet{Viani2017}.
Unfortunately, their attempt to calibrate the scaling relations is model dependent and also changes when adopting another source of spectroscopic information. Though they demonstrate that they found frequency-derived masses and radii in good agreement with dynamical results for four out of five giants in eclipsing binaries, the values that one obtains from their scaling relations using $\Delta \nu$ and $\nu _{\mathrm{max}}$ from \citet{Gaulme2016} are too small. This indicates that their average seismic parameters are systematically different, which complicates comparisons.  

Eclipsing binaries are the only stars for which precise, accurate, and model-independent radii as well as masses can be measured. Aided by modern observational techniques and analysis methods, these objects continue to provide the most stringent tests of stellar evolution theory and the asteroseismic methods. \citet{Brogaard2016} describe how eclipsing binary stars with an oscillating giant component can be used for establishing the accuracy level of masses and radii of giant stars measured with asteroseismic methods. Studies of a few individual systems were carried out by \citet{Frandsen2013}, \citet{Rawls2016}, and \citet{Themessl2018}. \citet{Gaulme2016} have published measurements of a larger sample of such eclipsing systems, three of which were re-measured by \citet{Brogaard2018}. \citet{Benbakoura2021} have expanded the known sample of useful SB2 systems by three, but with a limited measurement accuracy. The main aim of this paper is a detailed and accurate characterisation of KIC\,4054905, one of the systems measured by \citet{Benbakoura2021}.

In Sects.~\ref{sec:target} --~\ref{sec:parallax}, we present observations and precise and accurate measurements of masses, radii, effective temperatures, and metallicities of the giant stars in the eclipsing binary system KIC\,4054905. In Sect.~\ref{sec:seis} we derive asteroseismic parameters and make a first interpretation. Then, in Sect.~\ref{sec:compare} we firmly establish the age and evolutionary state of KIC\,4054905 before
comparing dynamical and asteroseismic parameters using scaling relations and grid modelling to establish the consistency level at this metallicity. Finally, in Sect.~\ref{sec:conclusion}, we summarise, conclude, and outline future work crucial to establishing and improving the accuracy of the asteroseismology of giant stars across all masses and metallicities.


\section{Target}
\label{sec:target}

Our target, KIC\,4054905, is listed by \citet{Gaulme2019} as an SB2 system with an oscillating giant component. It was analysed by \citet{Ou2019} from {\it Kepler} photometry alone, while \citet{Benbakoura2021} used multi-epoch spectroscopy in combination with {\it Kepler} photometry to establish dynamical parameters of both components and asteroseismic parameters for the oscillating giant. The precision of the dynamical parameters in that study was not high enough for detailed comparisons with asteroseismology. Furthermore, they claim the oscillating giant to be in the core-helium burning red clump (RC) phase of evolution based on asteroseismology, though this is at odds with other indicators. 

\section{Observations and observables}

For our analysis we employed time-series light curve observations and multi-epoch radial velocity (RV) observations as detailed below.

\subsection{\textit{Kepler} photometry}
\label{sec:phot}
We obtained photometric light curves of KIC\,4054905 in two ways. The \textit{Kepler} light curve for the eclipsing binary modelling was obtained by downloading light curve data from the MAST archive and extracting the PDCSAP\_FLUX. The reason why we chose the PDCSAP\_FLUX is because this has been corrected for light contamination from other stars, known as third light in binary star analysis. This light curve was normalised by fitting and dividing it by a second order polynomial to regions close to, but outside eclipses, for every single eclipse separately. Only photometry in, or very close to, eclipses was retained, and the flux was converted to relative magnitude with a mean of 0 mag outside eclipses. Uncertainties were set to the same number for all photometric measurements, which were first calculated as the root-mean-square (RMS) of regions outside of an eclipse, but afterwards re-scaled upwards in order for JKTEBOP \citep{Southworth2004} to produce a $\chi ^2$ = 1 for the light curve (cf. Sect.~\ref{sec:binary}). 

For our asteroseismic analysis, a \textit{Kepler} light curve was constructed as described in \citet{Brogaard2018}. Briefly, light curves were constructed from \textit{Kepler} pixel data (Jenkins et al. 2010) downloaded from the KASOC database\footnote{\url{kasoc.phys.au.dk}} using the procedure developed by S. Bloemen (private comm.) to automatically define pixel masks for aperture photometry. This light curve was corrected for jumps between observing quarters and concatenated. It was then median filtered using two filters of different widths, so as to account for both spurious and secular variations, with the final filter being a weighted sum of the two filters based on the variability in the light curve. In addition to the median filters, the signal from the eclipses was estimated and included in the final filter from the construction of the eclipse phase curve. This filtering allows one to isolate the different components of the light curve, and select which to be retained in the final light curve -- we refer interested readers to \citet{Handberg2014} for more details on the filtering methodology.

\subsection{Spectroscopy}

For spectroscopic follow-up observations, we used the FIES spectrograph at the Nordic Optical Telescope located at the Observatorio del Roque de los Muchachos on La Palma. The FIES spectra were obtained using the HIGHRES setting which results in a resolving power of $R\sim67000$.

Eleven epochs of observations were gathered between 06 April 2021 and 16 September 2021, each with an integration time of 3060 seconds
and with ThAr calibration frames taken immediately before each science observation. 

\subsubsection{Radial velocities}

To measure the RVs of the binary components at each epoch and to separate their spectra, we used the procedure described by \citet{Brogaard2018}. Briefly, the method combines a spectral separation code, closely following the description of \citet{Gonzalez2006} with the broadening function (BF) formalism by \citet{Rucinski1999,Rucinski2002} employing synthetic spectra from the grid of \citet{Coelho2005}. Four wavelength ranges were treated separately ($\lambda =$ 4500--5000, 5000--5500, 5500--5880, and 6000--6500 $\AA$). The gap between the last two wavelength ranges was introduced to avoid the region of the interstellar Na lines that causes problems for the spectral separation. For each epoch the final RV was taken as the mean of the results from each wavelength range and the RMS scatter across the wavelength was considered to be the uncertainty. Tables including the individual RV measurements of the components and the specific barycentric Julian dates, calculated using the software by \citet{Eastman2010}, are given in the appendix.

\subsubsection{Spectroscopic parameters}
\label{sec:specanal}
A classical equivalent-width spectral analysis was performed on the disentangled spectra, re-normalised according to the light ratio from the eclipsing binary analysis in Sect.~\ref{sec:binary}. The small variation of the light ratio with the wavelength was disregarded due to the following: We use spectral lines in the wavelength interval $5058.5$ {\bf --} $6786.9 \AA$. In this interval, the light ratio varies only between 0.20 at the blue end to 0.18 at the red end, if one assumes the spectroscopic $T_{\rm eff}$ of the primary, and the ratio of radii and the $K_P$ light ratio from the binary analysis. This also gives an effective temperature of the secondary of 5181 K, which is consistent with the spectroscopic $T_{\rm eff}$ well within the errorbar. Errors in the light ratio due to the neglect of its variation with wavelength are therefore only at the $\pm0.01$ level at the extreme ends, and thus insignificant compared to other error sources. In addition, due to the uncertainty of the derived effective temperatures, the uncertainty in the variation of the light ratio with wavelength is expected to be at a similar level. 
We determined $T_\text{eff}$, $v_\text{mic}$, [Fe/H], and [$\alpha$/Fe] for the two components following the suggested method (labelled DAOSPEC+C14) outlined in \citet{Slumstrup2019}. The line list used here consists of the available lines from Table A.1 of that paper. The $\log g$ values were fixed at the more precise estimates from the binary solution, 2.57 for the primary component and 3.44 for the secondary component. These values were determined using the masses and radii from Table 1 with  $\log g =$ 4.438 + $\log (M/R^2)$ with $M$ and $R$ in solar units. The auxiliary programme Abundance with SPECTRUM \citep{Gray1994} was used to determine the atmospheric parameters while assuming LTE, using MARCS stellar atmosphere models \citep{Gustafsson2008} and solar abundances from \citet{Grevesse1998}. We used astrophysical oscillator strengths, where the $\log gf$ values of each absorption line had been calibrated on a solar spectrum by adjusting until the well-established solar abundances were achieved \citep[for a further explanation, see][]{Slumstrup2019}. We report [$\alpha$/Fe], which in this paper is defined as $\frac{1}{4} \cdot \left( \text{ [Ca/Fe] + [Si/Fe] + [Mg/Fe] + [Ti/Fe] } \right)$, which yielded the parameters given in Table~\ref{table:specpar}. 

Due to the significantly lower contribution from the secondary component to the total flux, the uncertainties on the atmospheric parameters for this star are correspondingly higher. However, as seen, [Fe/H] and $[\rm \alpha/Fe]$ values are in very good agreement between components. Due to the much larger uncertainties on the parameters of the secondary, we adopted [Fe/H] and $[\rm \alpha/Fe]$ from the primary component for the continued analysis. We also calculated  

\begin{equation}
\rm [M/H]=[Fe/H] + \log ( 0.694 \times 10^{\rm [ \alpha/Fe]} + 0.306 ) = -0.35
\end{equation}
from \citet{Salaris1993}, a value of the overall metallicity of the components. This was used later for limb darkening calculations and isochrone comparisons.

The relatively low [Fe/H] and the high $\alpha$ enhancement indicates a thick disk star. We calculated (U, V, and W) Galactic velocity components using TOPCAT \citep{Taylor2019} with Gaia EDR3 data. For the RV component, we used our own system velocity measurement, since the Gaia RV value depends on the exact epochs of observation. The derived (U, V, and W) values are given in Table~\ref{tab:EBdata}, and also suggest a thick disk origin for KIC\,4054905.

\begin{table}
\caption{Spectroscopic parameters of KIC\,4054905.}             
\label{table:specpar}      
\centering                          
\begin{tabular}{l c c}        
\hline\hline                 
Parameter & Primary & Secondary \\    
\hline                        
$T_{\rm eff}$ (K) & $4850\pm70$ & $5260\pm240$ \\
$v_\text{mic}$ & $1.25\pm0.11$ & $1.8\pm0.8$ \\
$[\rm Fe/H]$ & $-0.60\pm0.02$ & $-0.55\pm0.16$ \\
$[\rm \alpha /Fe]$ & $+0.33\pm0.09$ & $+0.32\pm0.27$ \\
\hline                                   
\end{tabular}
\end{table}

\section{Eclipsing binary analysis}

\label{sec:binary}
To determine dynamical stellar parameters, we used the JKTEBOP eclipsing binary code \citep{Southworth2004} which is based on the EBOP programme developed by P. Etzel \citep{Etzel1981,Popper1981}. We made use of quadratic limb darkening \citep{Southworth2007}, the simultaneous fitting of the light curve and the measured RVs \citep{Southworth2013}, and numerical integration \citep{Southworth2011}, which is needed due to the long integration time of {\it Kepler}-long cadence photometry (24.9 minutes).

The input was our {\it Kepler} PDCSAP\_FLUX light curve and our RV measurements of KIC\,4054905. First guesses for the parameters were adopted from the analysis of \citet{Benbakoura2021}. The binary system contains two lower red giant branch (RGB) stars, and we refer to the largest one as the primary and the smaller one as the secondary star. 

We fitted for the following parameters: orbital period $P$, first eclipse of the primary component $T_{\rm p}$, central surface brightness ratio $J$, sum of the relative radii $r_{\rm p}+r_{\rm s}$, ratio of the radii $k=\frac{r_{\rm s}}{r_{\rm p}}$, orbit inclination $i$, eccentricity $e$, longitude of periastron $\omega$, RV semi-amplitudes of the components $K_{\rm p}$ and $K_{\rm s}$, and system velocities of the components $\gamma_{\rm p}$ and $\gamma_{\rm s}$. We allowed for two system velocities because the components and their analysis are affected by gravitational redshift \citep{Einstein1952} and convective blueshift \citep{Gray2009} effects differently. An after-the-fact evaluation of the derived system velocities showed a difference in accordance with expectations for both size and direction of the relative shift.

We used a quadratic limb darkening law with coefficients calculated using JKTLD \citep{Southworth2015} with tabulations for the $K_p$ bandpass by \citet{Sing2010}. We first used $T_{\rm eff}$ and [Fe/H] from \citet{Benbakoura2021} with log$g$ fixed from a preliminary binary solution. Then, with the light ratio determined from the binary analysis, we separated and re-normalised the spectra and determined the spectroscopic $T_{\rm eff}$ and [Fe/H] values as described in Sect.~\ref{sec:specanal} and used these in the subsequent iteration. New limb darkening coefficients were then calculated with JKTLD using these $T_{\rm eff}$ and log$g$ values to be used in the next JKTEBOP solution. We used the smallest possible [M/H]$=-0.3$ value in the \citet{Sing2010} tabulation, which is slightly larger than the value of [M/H]$=-0.35$ from our spectroscopic measurements. Therefore, we held the linear coefficients fixed and fitted for the non-linear coefficients in our final solution. We also carried out tests, instead fixing the non-linear coefficients and fitting the linear ones. This resulted in a primary radius smaller by 0.4\% = $1\sigma$. Fixing both the linear and non-linear coefficients gave a primary radius smaller by 0.9\% = $2.7\sigma$. Thus, limb darkening uncertainties are perhaps the dominating contributor to the radius uncertainty of the primary star, which is potentially as much as 0.9\% = $2.7\sigma$ smaller than our best fit, if fixed theoretical limb darkening coefficients are to be trusted more than fitting for one of them. This should be kept in mind when using the dynamical parameters as ground truth when comparing dynamical measurements to asteroseismic results.
Effects on other parameters were less significant, for example the maximum change to the secondary radius was at the level of 0.1\% = $0.3\sigma$.

Gravity darkening coefficients and reflection effects were set to zero due to our pre-analysis manipulation of the light curve. We tested that large changes to these numbers had negligible effects, as expected for nearly spherical stars.

For analyses of eclipsing binary stars, the so-called third light ($\ell_3$) is the fraction of light that does not originate from either of the two stars in the binary system. For \textit{Kepler} light curves, there is often a contribution to $\ell_3$ from nearby stars due to the pixel size corresponding to about 4x4 arcseconds. In addition to that, there could also be third light from an unresolved third component or a very close chance alignment.  

To set a limit on third light from an unresolved component in the system or an unresolved chance alignment, we investigated the spectra in more detail. From each of the 11 observed spectra, we subtracted the disentangled component spectra at their measured velocities. 
We calculated the BFs of the remainder to see if any additional signal appeared. We averaged these BFs across all observations,
since we expect a constant RV for a potential signal because our measured system velocity is stable over the time span of the observations.
By adding artificial spectra with a varying light ratio into the analysis, and comparing the BF peak amplitude to the maximum peak without
artificial signal, we found an upper limit on the light ratio of a potential third component with a similar effective temperature and rotational velocity as the A and B components of about 1\%. For a potential hotter third star and/or faster rotation, a larger light ratio will be possible. However, the largest true BF peak is located close to
zero RV, and is thus more likely to be caused by scattered sunlight than a signal from a third component. We thus find it reasonable to assume a maximum third light from an unresolved component of 1\%. If a 1\% third light is included in the binary analysis, both the masses decrease by just below 0.1\% (much less than $1\sigma$) and the radius is reduced by 0.4\% (about $1\sigma$) for the primary and is increased by 0.5\% (about $1\sigma$) for the secondary.

For the PDCSAP\_FLUX, the light contamination from known nearby stars has already been removed as part of the {\it Kepler} PDC data reduction. Therefore, third light ($\ell_3$ in JKTEBOP) was set to zero in the binary analysis, also assuming no unresolved companion in the system.

As a test, we also performed a model fit using the KASOC light curve (cf. Sect.~\ref{sec:phot}), where light contamination from other stars was treated as third light and was fixed to the mean contamination of quarters with an eclipse ($\ell_3=0.0129$) found on the \textit{Kepler} MAST webpage\footnote{https://archive.stsci.edu/kepler/data\_search/search.php}. 
That solution yielded identical masses, but a primary radius larger by 1.2\% and a secondary radius smaller by 1.2\%. This was indeed expected since the KASOC light curve uses a larger aperture than is used to estimate the contamination on MAST. Thus, the  contamination in the KASOC light curve is likely larger than $\ell_3=0.0129$. Tests showed that adopting $\ell_3=0.045$ yielded radii identical to the PDCSAP\_FLUX solution. Although the level of contamination is relatively large for KIC\,4054905, this should serve as a reminder to carefully consider the contribution of third light when analysing similar systems in the future. Either the PDCSAP\_FLUX light curve must be used, or contamination must be calculated for the specific aperture used. 

\begin{figure}
   \centering
   \includegraphics[width=8.6cm]{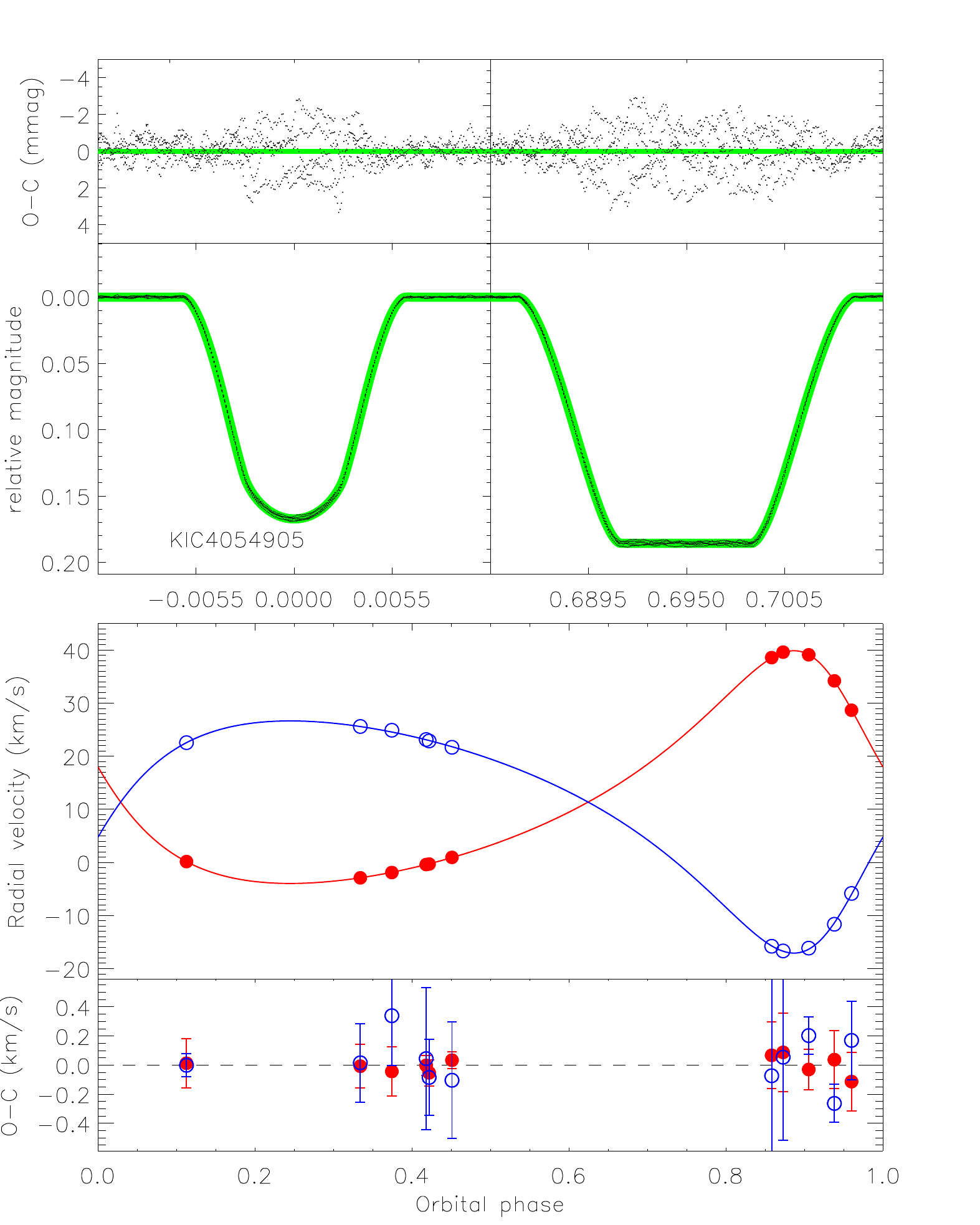}
   \caption{Binary model fit to the {\it Kepler} light curve (upper panels) and RVs (lower panels) for KIC\,4054905. Red indicates the primary component, and blue is for the secondary component. Filled and open circles represent our RV measurements of the primary and secondary component, respectively. }
             \label{fig:4054905_model}%
    \end{figure}

The optimal JKTEBOP solution using the PDCSAP\_FLUX light curve is compared to the observed PDCSAP\_FLUX light curve and measured RVs in Fig.~\ref{fig:4054905_model}. It is clear from the light curve O-C diagram that the residuals are dominated by the solar-like oscillations rather than random errors. We therefore used the residual-permutation uncertainty estimation method of JKTEBOP (task 9), which accounts for correlated noise when estimating parameter uncertainties. The final parameters and their uncertainties are given in Table~\ref{tab:EBdata}.

Our mass and radius estimates are close to the corresponding measurements by \citet{Benbakoura2021}, but much more precise. Specifically, using their uncertainties, both their measured masses agree with ours well within $1\sigma$, and radii agree within $2.1\sigma$ for the primary and $0.67\sigma$ for the secondary component. Using our uncertainties instead, the mass values agree within $0.5\sigma$ for the primary, but $7.9\sigma$ for the secondary; whereas, the radii differ by $6.4\sigma$ and $2.0\sigma$. These differences are mainly due to our significantly more precise RV measurements, though they are also due to a more precise treatment of third light and limb darkening.

\begin{figure}
   \centering
   \includegraphics[width=8.6cm]{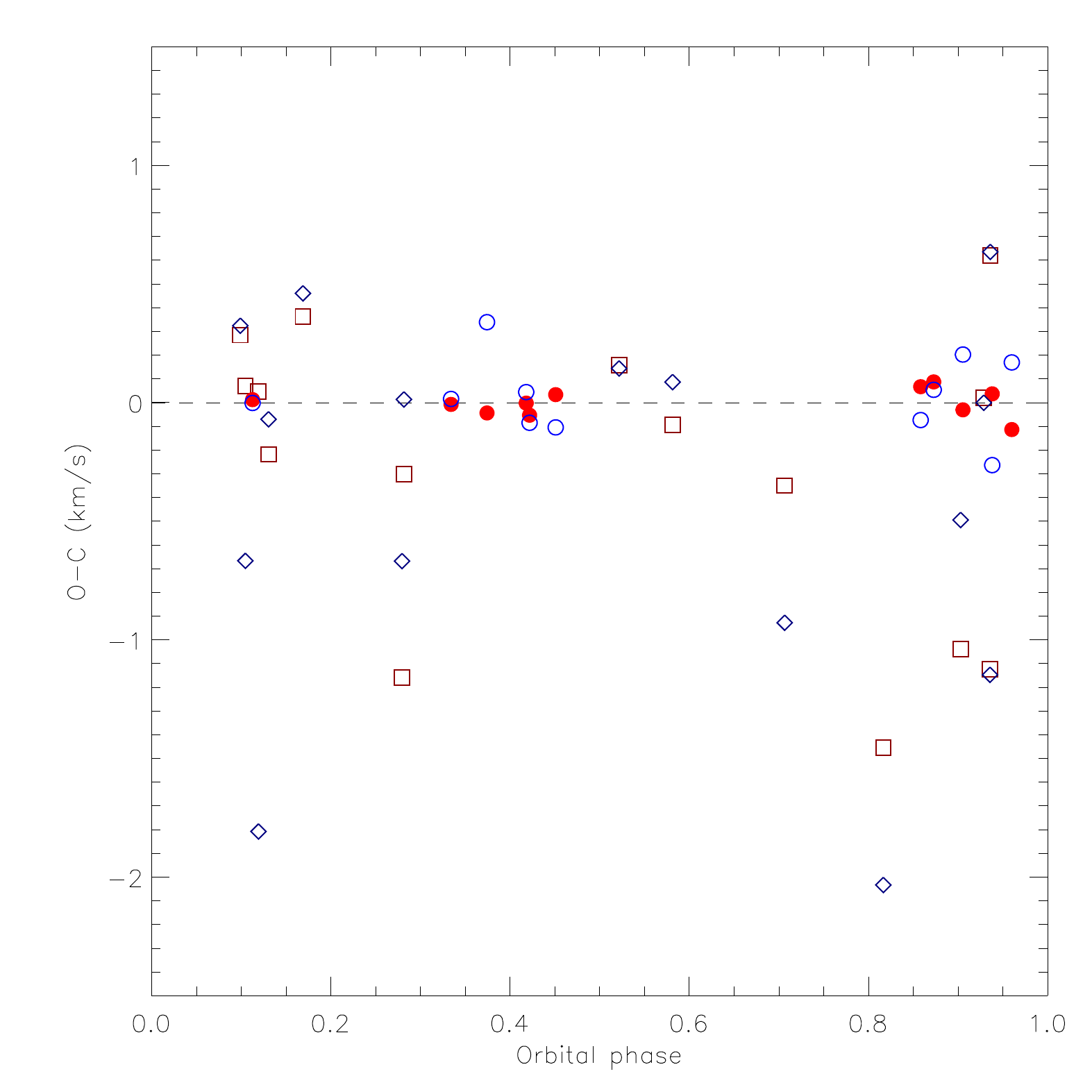}
   \caption{O-C diagram from the binary model in Fig.~\ref{fig:4054905_model}. Red solid circles correspond to our measurements of the primary component, blue open circles are for the secondary component. Dark red squares and navy diamonds represent the measurements of \citet{Benbakoura2021} for the primary and secondary star, respectively, with phases and O-C calculated relative to our binary solution.}
             \label{fig:4054905_model_OC}%
    \end{figure}

The RV O-C diagrams in Fig.~\ref{fig:4054905_model_OC} show our residuals as well as those of \citet{Benbakoura2021} when their RV measurements are phased to our dynamical solution. In general, our results are much more precise, which is not surprising given the higher resolution of our spectra. However, the comparison strongly indicates that the measurements by \citet{Benbakoura2021} suffer from epoch-to-epoch RV zero-point issues because the (O-C) values of their measurements are much larger than the claimed RV uncertainties and with an offset direction correlated between the two components for almost all epochs of observation. 

\begin{table}
\centering
\caption{Properties of the eclipsing binary KIC4054905.}
\label{tab:EBdata}
\begin{tabular}{lr}
\hline
\hline
RA  (J2000)\tablefootmark{a}                                &  $19:23:40.543$   \\
DEC (J2000)\tablefootmark{a}                                &  $+39:10 :11.60$\\
$K_p$\tablefootmark{a}                                              & 12.984    \\ 
\hline
\it{Dynamical parameters:} & \\
Orbital period (days)                               & $274.72881(97)$   \\
Reference time $\rm T_{\rm p}$                     & $55503.4588(16)$  \\
Inclination $i$ ($\circ$)                           & $89.284(13)$       \\
Eccentricity $e$                           & $0.37201(47)$      \\
Periastron longitude $\omega$ ($\circ$)                       & $35.17(12)$\\
Sum of the fractional radii $r_{\rm s}+r_{\rm p}$ & $0.051903(90)$\\
Ratio of the radii $k$                              & $0.36960(58)$\\
Surface brightness ratio $J$                        & $1.302(18)$\\
$\frac{L_{\rm s}}{L_{\rm p}} | K_p$                 & $0.18631(16)$\\
$K_{\rm p}$  $\rm (km\cdot s^{-1})$                & $21.915(34)$ \\
$K_{\rm s}$  $\rm (km\cdot s^{-1})$                                       & $21.87(13)$  \\
semi-major axis $a (\rm R_{\odot})$                   & $260.69(72)$\\
$\gamma_{\rm p}$ $\rm (km\cdot s^{-1})$                                  & $11.295(24)$ \\
$\gamma_{\rm s}$ $\rm (km\cdot s^{-1})$                                  & $11.43(11)$  \\
Mass$_{\rm p}(M_{\odot})$                          & $0.9544(94)$\\
Mass$_{\rm s}(M_{\odot})$                          & $0.9566(59)$\\
Radius$_{\rm p}(R_{\odot})$                        & $8.364(27)$     \\
Radius$_{\rm s}(R_{\odot})$                        & $3.0911(94)$      \\
log$g_{\rm p}$ (cgs)                               & $2.5730(26)$      \\
log$g_{\rm s}$ (cgs)                               & $3.4385(19)$ \\
$\rho_{\rm p} (\rho_{\odot}$) & 0.001631(23)\\
$L_{\rm p}(L_{\odot})$ from $R$ and $T_{\rm eff}$  & $34.87(2.02)$\\
$L_{\rm s}(L_{\odot})$ from $R$ and $T_{\rm eff}$ & $6.59(1.21)$\\ 
\hline
\it{Gaia parameters:} & \\
$G$ (mag) & 13.011\\
$\frac{L_{\rm s}}{L_{\rm p}} | G$  & $0.18735$\\
$G_{\rm p}$ (mag) & 13.197\\
$G_{\rm s}$ (mag) & 15.158\\
$A_G$ (mag) & 0.291\\
$\varpi$ (mas) & 0.42139(1145)\\
$\varpi_{\rm zp cor, Lindegren}$ (mas)\tablefootmark{b} & $-0.02189$\\
$\nu _{\rm eff}$ & 1.4485245\\
ecl\_lat & 60.1765086\\
$BC_{G,\rm p}$ (mag) & $-0.070$\\
$BC_{G,\rm s}$ (mag) & $+0.004$\\
$L_{\rm p}(L_{\odot})$ from $G$\tablefootmark{c}  & $29.47(2.13)$\\
$L_{\rm s}(L_{\odot})$ from $G$\tablefootmark{c} & $5.16(37)$\\ 
$L_{\rm p}(L_{\odot})$ from $K_S$\tablefootmark{c}  & $32.65(1.70)$\\
$L_{\rm s}(L_{\odot})$ from $K_S$\tablefootmark{c} & $6.15(32)$\\ 
U$\rm (km\cdot s^{-1})$\tablefootmark{d} & $139.5$\\
V$\rm (km\cdot s^{-1})$\tablefootmark{d} & $-28.5$\\
W$\rm (km\cdot s^{-1})$\tablefootmark{d} & $-29.6$\\
\hline
\it{Spectroscopic parameters:} & \\
$T_{\rm eff,p}$ (K)                                    & $4850(70)$\\
$T_{\rm eff,s}$ (K)                                      & $5260(240)$   \\
$\rm [Fe/H]$                                        & $-0.60(2)$  \\
$\rm [\alpha /Fe]$                                          & $+0.33(09)$  \\
\hline
\it{Asteroseismic stellar parameters:} & \\
$f_{\Delta\nu}$ correction factor\tablefootmark{e}  & 0.9644 \\    
Mass$_{\rm p, scaling}(M_{\odot})$ & 1.024(48)\\
Radius$_{\rm p, scaling}(R_{\odot})$ & 8.44(15)\\
log$g_{\rm p, scaling}$ (cgs)  & 2.5955(58) \\
$\rho_{\rm p, scaling} (\rho_{\odot})$ & 0.001701(21)\\
\hline
\it{Age estimate:} & \\
Age$_{\rm dyn}$ (Gyr)           & $9.9\pm0.6$            \\
\hline
\end{tabular}

\tablefoot{
\scriptsize
Solar units are from \citet{IAU2015}.
\tablefoottext{a}{From the KIC.}
\tablefoottext{b}{Gaia EDR3 correction to the parallax using \citet{Lindegren2021}. To be subtracted from $\varpi$ in the line above.}
\tablefoottext{c}{From Gaia EDR3 and DR2 parameters.}
\tablefoottext{d}{From the Gaia EDR3 catalogue using TOPCAT.}
\tablefoottext{e}{Using \citet{Rodrigues2017}.}
}
\end{table}

   \begin{figure}
   \centering
   \includegraphics[width=8.6cm]{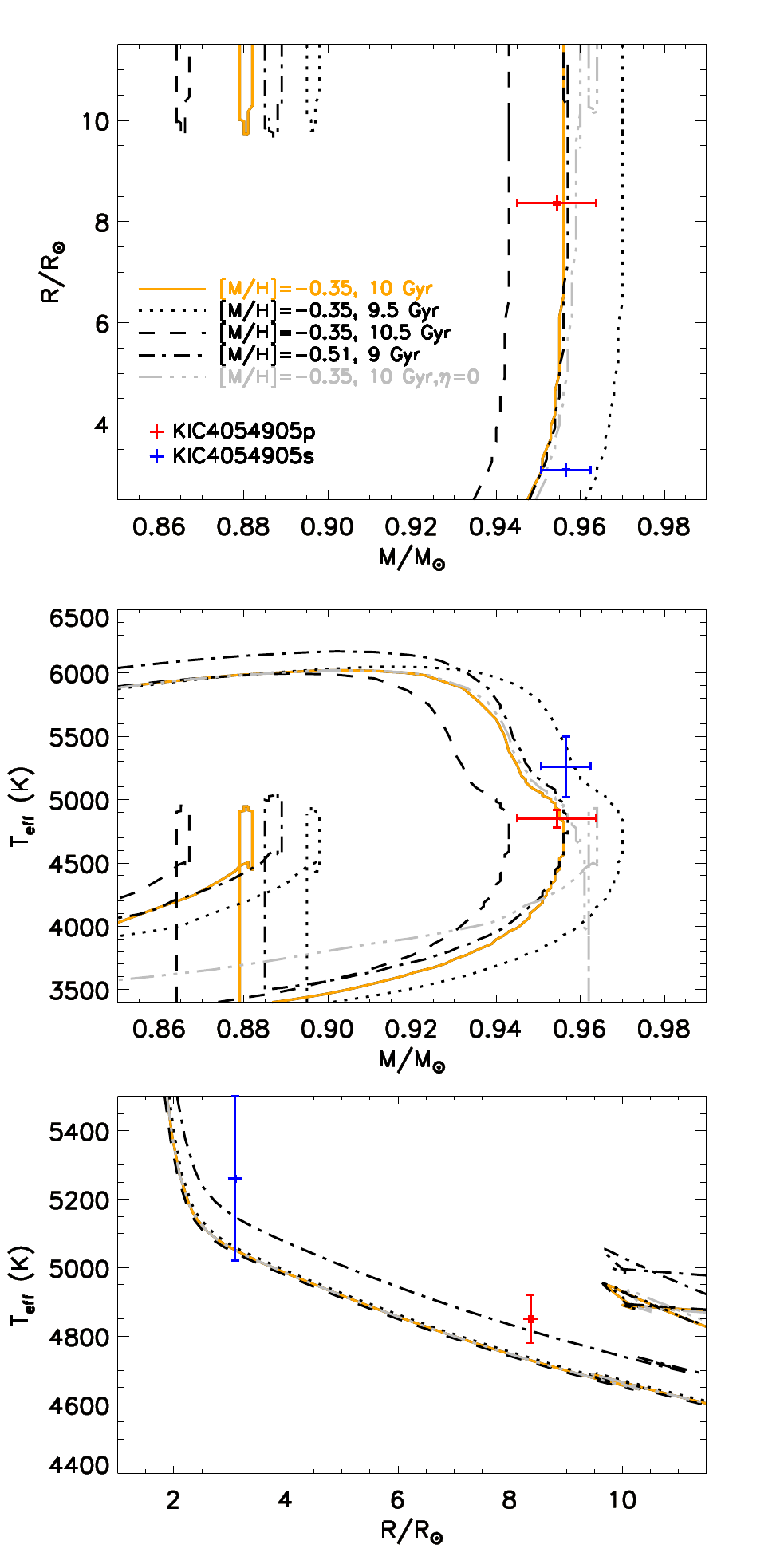}
   \caption{Mass-radius, mass-$T_{\rm eff}$, and radius-$T_{\rm eff}$ diagrams for KIC\,4054905 compared to PARSEC isochrones. Red symbols indicate the most evolved component, while blue symbols indicate the least evolved one.}
             \label{fig:mrt}%
    \end{figure}

%
%

\section{Luminosity from the Gaia parallax}
\label{sec:parallax}
The luminosity of both binary components can be estimated using their parallax which was measured by the Gaia mission \citep{Gaia2016}. The parallax of KIC4054905 was extracted from the Gaia EDR3 catalogue \citep{Gaia2021} along with an apparent magnitude in the Gaia $G$ band from the DR2 catalogue \citep{Gaia2018}. Interstellar absorption, $A_G=0.291$, was inferred from $E(B-V)=0.0106\pm0.02$, which was obtained from the reddening map of \citet{Green2019}. The parallax was zero-point corrected by using the python code provided by \citet{Lindegren2021}. We then separated the magnitudes of the binary components using the light ratio from the dynamical model, after it was converted from the \textit{Kepler} band to the Gaia $G$ band using Planck functions and the spectroscopic $T_{\rm eff}$ values.

To obtain the absolute magnitudes, we used equation (10) of \citet{Torres2010} generalised to the Gaia $G$-band: 
\begin{equation}\label{eq:09}
M_{G \rm p,s} = -2.5\log\left(\frac{L_{bol \rm p,s}}{L_{\odot}}\right)+V_{\odot}+31.572-BC_{G \rm p,s}+BC_{V,\odot}. 
\end{equation}
Here, $V_{\odot} = -26.76$, as recommended by \citet{Torres2010}, and $BC_{V, \odot} = -0.068$, as obtained from the calibration of \citet{Casagrande2014}. We note that p and s represent the primary and secondary component, respectively. The bolometric corrections were obtained from \citet{Casagrande2014, Casagrande2018B, Casagrande2018A}. 
Combining Eqn.~(\ref{eq:09}) with the definition of the apparent distance modulus, we derived equation~(\ref{eq:10}) from which we obtained the luminosity of each binary component, p or s, respectively: 
\begin{eqnarray}\label{eq:10}
\frac{L_{\rm p,s}}{L_{\odot}} & = 10^{\left(\frac{5\log\left(\frac{1000 \rm{mas}}{\varpi}\right)-0.256+A_G-G_{\rm p,s}-BC_{G \rm p,s}}{2.5}\right)}. 
\end{eqnarray}

We also repeated the procedure for the $K_S$ band, which is less sensitive to reddening. These estimates for $L$ from the parallax and $G$ and $K_S$ photometry are given in Table~\ref{tab:EBdata} in the section with Gaia parameters. The corresponding numbers calculated from the dynamical radii and spectroscopic effective temperatures are given in the same table in the section with dynamical parameters. The uncertainties are large enough so that the rather different numbers for the luminosities cannot be taken as evidence that something is wrong; however, they do indicate a tension. Tests showed that if we adopted $T_{\rm eff}=4700$\,K for the primary component instead of the spectroscopic value of $T_{\rm eff}=4850$\,K, then the three luminosity estimates for the primary are in much better agreement:
$L_p/L_{\odot}=30.75\pm1.88, 30.64\pm2.22$, and $29.91\pm1.58$ from $T_{\rm eff}$, $G$ band, and $K_S$ band, respectively, compared to
$L_p/L_{\odot}=34.87\pm2.01, 29.47\pm2.14$, and $ 32.65\pm1.71$ 
using the spectroscopic $T_{\rm eff}$. Interestingly, the $G-K_S$ and $G_{\rm BP}-K_S$ colours of the combined light along with the colour-temperature relations of \citet{Casagrande2021} also suggest $T_{\rm eff}$ just above 4700\,K (4723 K and 4749 K, respectively) when calculated assuming $E(B-V)=0.0106$, thus also supporting a lower $T_{\rm eff}$ than the spectroscopic one. 

We also estimated the potential systematic error in the Gaia parallax due to the orbital motion of KIC\,4054905 using the method outlined in appendix C of \citet{Rappaport2022}, to which we refer the reader for details. Briefly, we first calculated the angular size of the semi-major axis of the centre of light orbit at the parallax distance, a(col). We then used the orbital parameters to determine the position vector of the centre of light at as many times as observed by Gaia DR3 (46), assuming that the observations were equally spaced within the time span of Gaia DR3. Using this information, we determined the RMS of the projections onto the plane of the sky to get an idea of the size of the deviations, RMS(col). Finally, we arrived at our crude estimate of the potential systematic error in the Gaia parallax due to the orbital motion, error(col), by dividing RMS(col) by the number of Gaia observations, so as to allow the potential systematic error to shrink with the number of measurements. We give the numbers in Table~\ref{table:astpar} along with specific Gaia EDR3 parameters for comparison. This shows that the astrometric orbit of the centre of light is relatively large compared to the parallax angle. The potential systematic error(col) is three times larger than the Gaia parallax error, though still only half of the astrometric\_excess\_noise. Thus a significant systematic error could be affecting our results based on the Gaia parallax.

\begin{table}
\caption{Astrometric parameters of KIC\,4054905.}             
\label{table:astpar}      
\centering                          
\begin{tabular}{l c}        
\hline\hline                 
Parameter & value \\    
\hline                        
Gaia EDR3 values: & \\
astrometric\_excess\_noise $(\mu\rm as)$       & 62 \\
astrometric\_excess\_noise\_significance    & 3.94  \\
RUWE                            & 1.01  \\
parallax $(\mu\rm as)$                        & 421  \\
parallax error $(\mu as)$                 & 11\\
astrometric\_matched\_transits    & 46         \\
\hline
This work: & \\
a(col) $(\mu\rm as)$                          & 175         \\
RMS(col) $(\mu\rm as)$                        & 157         \\
error(col) $(\mu\rm as)$                      & 35         \\

\hline                                   
\end{tabular}
\end{table}


We decided to adopt the spectroscopic $T_{\rm eff}$ because of these parallax issues which are further complicated by the parallax zero-point correction, which depends on both the magnitude and colour of single stars, while we use numbers for the combined light of a binary.
In addition to that, the $T_{\rm eff}$ uncertainties due to reddening uncertainties are fairly large. Also, when we tried to fix $T_{\rm eff}=4700$\,K in the spectroscopic analysis, we obtained [Fe/H]$=-0.54$ and $[\alpha/\rm Fe]=+0.30$, which seems to be an unusual abundance combination. While this cannot be ruled out, it could be taken as an indication that $T_{\rm eff}=4700$\,K is too low. In order to understand these issues better, we need more spectroscopic observations to increase signal-to-noise in the spectroscopic analysis and/or later releases of Gaia data, which may include a binary orbit and therefore a more precise parallax measurement. We note in this respect that even in the full Gaia DR3 release, KIC\,4054905 is not listed as a non-single object; although, it is a known eclipsing binary.


\section{Asteroseismic analysis}
\label{sec:seis}

The primary component of the eclipsing binary displays solar-like oscillations. We performed an asteroseismic analysis following the procedures described in \citet{Arentoft2017} and \citet{Brogaard2021}. The asteroseimic parameters derived are given in Table~\ref{tab:seis} for reference, though we only make use of $\nu_{\text{max}}$, $\Delta \nu_0$, and $\Delta P_{\rm obs}$ in the following.

We note that $\nu_{\text{max}}$ was determined with two different background models, eqn. (4) and (5) from \citet{Handberg2017} denoted by H4 and H5, as in \citet{Thomsen2022}, who found a very significant $\nu_{\text{max}}$ dependence on the background for the KIC\,8430105 giant+main sequence binary. Since we did not find a significant difference for KIC\,4054905A, we adopted $\nu_{\text{max}}=48.44\pm0.55\,\mu$Hz which was determined with H5 to be consistent with our other works.

\begin{table}
\centering
\caption{Asteroseismic parameters for KIC4054905A.}
\begin{tabular}{lc}
\hline
\noalign{\smallskip}
Parameter & Value \\
\noalign{\smallskip} \hline
$\Delta\nu_{\rm ps}$\tablefootmark{a} ($\mu$Hz) & 5.404$\pm$0.150\\
$\Delta\nu_{\rm 0}$\tablefootmark{b} ($\mu$Hz) & 5.366$\pm$0.033\\
$\delta_{\rm 02}$ ($\mu$Hz) & 0.744$\pm$0.061 \\ 
$\epsilon$ & 1.115$\pm$0.051\\
$\nu_{\rm max}$ ($\mu$Hz) (H5 background)& 48.44$\pm$0.55  \\
$\nu_{\rm max}$ ($\mu$Hz) (H4 background)& 48.11$\pm$0.44   \\
$\Delta$P$_{\rm obs}$ (s) & 52.6$\pm$2.3\\
$\Delta\nu_{\rm c}$ ($\mu$Hz) & 5.390$\pm$0.020\\
$\epsilon_{\rm c}$ & 1.064$\pm$0.030\\
\hline
\noalign{\smallskip}
\end{tabular}

\tablefoot{
\scriptsize
\tablefoottext{a}{The large separation determined directly from the power spectrum.}\\
\tablefoottext{b}{The large separation determined from individual frequencies.}\\ 
}

\label{tab:seis}
\end{table}

\subsection{Asteroseismic stellar parameters}
Using the observed values of the asteroseismic parameters $\nu_{\text{max}}$ and $\Delta \nu_0$, Equations (\ref{eq:03})-(\ref{eq:04}) were used to obtain values of the mass and radius of the oscillating primary component. This yielded $M/M_\odot = 1.024(48)$ and $R/R_\odot = 8.44(15)$, applying the correction factors $f_{\nu_{\text{max}}} = 1$ and $f_{\Delta \nu} = 0.9644$. The latter was estimated using a correction to $\Delta \nu$ inferred graphically from fig. 3 in \citet{Rodrigues2017}. The RGB phase of evolution was assumed since $f_{\Delta \nu}$ for a RC assumption suggests $f_{\Delta \nu} > 1$, which would result in asteroseismic masses and radii much larger than the dynamical values. More evidence supporting the RGB evolutionary phase is provided in Sect.~\ref{sec:compare}. 


By using Equations (\ref{eq:01})-(\ref{eq:02}), the surface gravity and mean densities were determined to be $g/g_\odot = 0.01436(19)$ and $\rho/\rho_\odot = 0.001701(21)$. All the asteroseismic stellar parameters are given in Table~\ref{tab:EBdata}.

\subsection{Evolutionary state}

\citet{Benbakoura2021} found a value of the asymptotic period spacing of mixed modes $\Delta\Pi_{\rm 1,asym} = 159.5 s$ for the oscillating giant and used this to classify the star as belonging to the helium-burning RC phase with reference to \citet{Bedding2011}. This is, however, incorrect because the parameter that distinguishes the evolutionary phase in \citet{Bedding2011} is not the asymptotic period spacing $\Delta\Pi_{\rm 1,asym}$, but rather the mean observed period spacing, often denoted by $\Delta P_{\rm obs}$. Our measured value of $\Delta P_{\rm obs}=52.6\pm2.3 s$ puts the star clearly in the RGB phase of evolution according to \citet{Bedding2011}. This is also consistent with the evolutionary state determined for KIC\,4054905 by \citet{Elsworth2017}.
The asymptotic $\Delta\Pi_{\rm 1,asym}$ value would also be expected to be much larger than 159.5 s (and $\Delta \nu$ much smaller than measured) for the low measured dynamical mass if the star belonged to the RC \citep[figure 1]{Mosser2014}. Our own attempt to measure $\Delta\Pi_{\rm 1,asym}$ gave a value in rough agreement with \citet{Benbakoura2021}, but it was deemed an untrustworthy solution because all of the individual period spacings were far from the asymptotic level. Our asteroseismic classification as an RGB star is further supported by the model comparisons in section ~\ref{sec:compare}.

\section{Discussion}
\label{sec:compare}
 
\subsection{Model comparisons}

Fig.~\ref{fig:mrt} compares masses, radii, and effective temperatures of the KIC\,4054905 components to PARSEC \citep{Bressan2012} isochrones in various diagrams. The [M/H]$=-0.35$ value was chosen by using the measured [Fe/H]$=-0.60$ from the primary component with the formula from \citet{Salaris1993} to account for the measured alpha enhancement of [$\rm \alpha /Fe]=+0.33$. Instead, assuming [Fe/H]$=-0.72$ from \citet{Benbakoura2021} with [$\rm \alpha /Fe$]$=+0.25$, as inferred for thick disk stars by eye from the bottom panel of fig. 5 in \citet{Miglio2021}, leads to [M/H]$=-0.51$ alternatively. An isochrone with this metallicity is therefore included as well to evaluate the effects of composition uncertainty. 

The mass-radius diagram shows that both components match the RGB phase at an age close to 10 Gyr assuming a thick disk origin with high [$\rm \alpha /Fe]$. The uncertainty for a fixed composition in the mass-radius diagram can be inferred by comparing the size of the mass errorbars to the mass difference between the two isochrones with an identical composition. This yields a 1$\sigma$ uncertainty of only $\pm0.3$ Gyr from the primary or $\pm0.2$ Gyr from the secondary. Instead, demanding that both components are matched within 1$\sigma$ in mass would result in an even smaller uncertainty close to $\pm0.1$ Gyr.
Comparing the isochrones of a different age and composition shows that an uncertainty of $\pm0.11$ on [M/H] leads to an age uncertainty of about $\pm0.5$ Gyr. Thus, the dominant age uncertainty comes from the uncertainty in composition.
Similar comparisons to isochrones from various other stellar models that were investigated in \citet{Tayar2022} resulted in older ages by up to about 1 Gyr. Therefore, we decided to investigate the age dependence of the assumed helium content and the \citet{Salaris1993} approximation using Victoria isochrones \citep{VandenBerg2014} where the helium content and alpha enhancement can be chosen freely. This showed that the Salaris approximation yields ages that are only 0.1 Gyr too old compared to the corresponding alpha-enhanced composition, and thus it seems to be a good approximation. The helium dependence turned out to be the major reason for the age differences from the various models. Specifically, we obtained 
$\Delta\rm{age}$/$\Delta Y=-70$\,Gyr/(mass fraction) given that a reduction in $Y$ by 0.01 corresponded to a 0.7 Gyr older age. Since the PARSEC isochrones have the steepest helium enrichment law of all the models we investigated, this explains why these models predict the youngest age for KIC\,4054905.  

Our best age estimate is thus $9.9\pm0.6$ Gyr where the uncertainty reflects only the observational uncertainties in mass and composition. The systematic age uncertainty due to the assumed helium content is of a similar size.

\citet{Benbakoura2021} found that the oscillating giant is in the RC phase. Fig.~\ref{fig:mrt} shows that this cannot be true unless mass loss has been very close to zero and unless the radius of the clump phase is significantly smaller than inferred from current stellar models. Only then would an isochrone be able to match both giants simultaneously in the mass radius diagram; the grey isochrone assuming no mass loss ($\eta=0$) can produce similar masses for the components as observed, but the model radius of stars in the RC phase is much larger than observed for the oscillating giant. \citet{Li2022Nat} demonstrate that mass transfer in close binary systems can produce a giant in the RC phase with a radius smaller than for single star evolution, but only if the initial mass was larger than 1.8 $M_{\odot}$. Even if the primary component of KIC\,4054905 could have lost about half of its original mass, we would not have been able to identify a scenario that would simultaneously result in a secondary star with a low mass that is in the giant phase of evolution. Much more likely is the alternative scenario with two near-identical mass early RGB stars, which is a natural result of single-star evolution in a well-detached binary with a mass ratio close to one. We conclude from this that the oscillating giant is in the RGB phase of evolution.

The bottom panel of Fig.~\ref{fig:mrt} shows that the effective temperatures are not well-matched by the isochrones unless the metallicity is lower than measured. This could also be another indication that the spectroscopic temperatures are too high, as suggested by the Gaia parallax distance and the colours of KIC\,4054905, but the $T_{\rm eff}$ scale of stellar models is sensitive to adopted surface boundary conditions and the mixing length parameter, rendering this an uncertain indication.

\subsection{Asteroseismic comparisons for the oscillating giant}

We used the asteroseismic scaling relations with a correction to $\Delta \nu$ inferred graphically from fig. 3 in \citet{Rodrigues2017}. We assumed the RGB phase of evolution and the dynamical mass to obtain $f_{\rm \Delta \nu}=0.9644$ self-consistently between the two diagrams. We adopted $f_{\nu_{\rm{max}}}=1$. This yields $M=1.025\pm0.048 M_{\odot}$ and $R=8.45\pm0.15R_{\odot}$. The asteroseismic mass is $1.5\sigma = 7\%$ larger than the dynamical value, while the numbers are $0.46\sigma = 0.9\%$ for the radius. Both asteroseismic log$g$ and $\rho$ from Eqns. (1)-(2) are larger than the corresponding dynamical values, as can be seen by comparing the numbers in Table~\ref{tab:EBdata}.

If we had instead adopted the values and uncertainties for the asteroseismic mean parameters from \citet{Benbakoura2021}, we would have obtained $M=0.966\pm0.026 M_{\odot}$ and $R=8.226\pm0.076R_{\odot}$  instead. The mass would then be fully consistent with the dynamical value and the radius would be 1.8$\sigma$ lower than the dynamical value. However, while a solution in between ours and that of \citet{Benbakoura2021} would then seem plausible, it is likely more complicated than it appears since both log$g$ and $\rho$ remain larger than the dynamical values at log$g=2.593\pm$0.004 and $\frac{\rho}{\rho_{\odot}}=0.001735\pm0.000006$. 

It is worth recalling that if the helium-burning phase is assumed, then $f_{\rm \Delta \nu} > 1.0$, which in turn results in masses and radii much larger than the dynamical values regardless of which asteroseismic mean parameters are used. This further supports our other piece of evidence that the star is on the RGB.

According to figure 3 of \citet{Viani2017}, $f_{\nu_{\rm max}}$=1.01 is suitable for KIC\,4054905A. Adopting it instead of $f_{\nu_{\rm max}}$=1 would change our scaling relation results to $M=0.995\pm0.047 M_{\odot}$, $R=8.36\pm0.15R_{\odot}$, log$g=2.591\pm$0.006, and $\frac{\rho}{\rho_{\odot}}=0.001701\pm0.000021$. The mass and radius are then consistent, whereas this is  still not the case for log$g$ and $\rho$. While the $\nu_{\rm max}$ correction does constitute an improvement, more stars with similar quality data are needed to know whether this is something that should be implemented and whether it has the right dependence on stellar parameters. The empirical calibration by \citet{Li2022} suggests a stronger variation of $f_{\nu_{\rm max}}$ with metallicity than \citet{Viani2017}, but that could be dependent on the source and analysis of the spectroscopic data. From Eqn. (2), we obtain $f_{\nu_{\rm max}}=1.053$ if we adopt log$g$ from the dynamical analysis.  

We also tried using equations (15)-(18) from \citet{Handberg2017} to estimate the stellar mass, using the independent luminosity that we derived from Gaia data. These four equations yield consistent masses (with $f_{\nu_{\rm max}}=1.01$ and $f_{\rm \Delta \nu}=0.9644$) within less than $0.01 M_\odot$, but the mass is $0.99 M_\odot$, which is still larger than the dynamical mass measurement. Within the uncertainty of the luminosity, solutions exist where the mass from all equations are in agreement with -- and their mean is identical to -- the dynamical value. While equation-to-equation variation increases, differences to their mean remains within the $1\sigma$ uncertainty.

Recently, \citet{Joergensen2020} tested various surface correction formulations with model fitting of individual frequencies for a number of oscillating giant stars in eclipsing binaries including KIC\,4054905. Although the fitting also included either $\nu_{\rm max}$ or the dynamical radius, these fits yielded results that are in good agreement with our dynamical mass and radius. The exact values that they derived are difficult to extract given that they only provide plots, not tables; however, agreement seems to be within 1$\sigma$ uncertainties. This suggests that model fitting of individual frequencies is currently more accurate than using the scaling relations with corrections. This is further supported by the study of \citet{Li2022} who attempted to perform a calibration of the asteroseismic scaling relations based on the model fitting of individual frequencies including surface corrections using a large sample of single giant stars. For four out of five giants in known eclipsing systems, they obtained masses and radii in good agreement with dynamical values, though, suspiciously, the asteroseismic numbers were all slightly on the low side of the dynamical ones. (The fifth star was compared using different metallicities so it cannot be interpreted easily.)
We used the APOGEE calibration of \citet{Li2022} on KIC\,4054905A and obtained $M=0.958\pm0.044 M_{\odot}$ and $R=8.21\pm0.15R_{\odot}$, in excellent agreement with our dynamical measurements; however, their corrected scaling relations still give asteroseismic log$g$ and $\rho$ values larger than the corresponding dynamical ones, log$g=2.591\pm$0.006 and $\frac{\rho}{\rho_{\odot}}=0.001730\pm0.000020$. This suggests that their derived scaling relations might still not be quite right. In defence of \citet{Li2022}, KIC\,4054905 is slightly outside of the metallicity range of their calibration, which further depends on the source of spectroscopic information. We also remind the reader of the potential systematic error in the dynamical radii due to limb darkening which is, in this case, large enough to explain discrepancies between dynamical and asteroseismic log$g$ and $\frac{\rho}{\rho_{\odot}}$.

Because of the previously mentioned indications that our spectroscopic $T_{\rm eff}$s could be too hot, we also tried asteroseismic calculations assuming $T_{\rm eff}$=4700\,K and the corresponding luminosity using the $K_S$ band with the Gaia parallax. This resulted in lower mass and radius estimates consistent with the dynamical results within 1$\sigma$ uncertainties when using the \citet{Rodrigues2017} correction, and within less than 2$\sigma$ for the \citet{Li2022} correction. 
Our comparisons seem to indicate that consistency between dynamical and asteroseismic parameters of KIC\,4054905A can be obtained, and that there is potential for reaching a version of the asteroseismic scaling relations which is accurate to the precision level. However, increasing the sample of calibrators remains crucial to reach correct conclusions about dependencies. Accuracy cannot be tested beyond the asteroseismic precision level, which can only be improved by an ensemble analysis.

\section{Conclusions and outlook}
\label{sec:conclusion}

We measured precise dynamical properties of the eclipsing binary KIC\,4054905, reaching better than 1\% precision for mass, and 0.4\% for radius, though with a potential bias on the latter of up to $-1\%$ due to limb darkening uncertainties. From disentangled component spectra, we measured spectroscopic $T_{\rm eff}$ of the components, along with [Fe/H] and $[\alpha \rm/Fe]$. There were indications that the spectroscopic $T_{\rm eff}$ could be too high. 

\textit{Kepler} data were used to measure average asteroseismic parameters and consistency checks were done using data from Gaia.
From the combined analysis, KIC\,4054905 was firmly identified as two near-identical mass 10 Gyr old stars on the lower RGB belonging to the thick disk.
Comparisons between dynamical and asteroseismic parameters illustrates the current level of consistency and points towards potential issues and future solutions.

To keep moving forward in establishing the accuracy level of asteroseismic scaling relations and more detailed asteroseismic methods, it is crucial to continue increasing and improving the sample of stars with high-precision, independent-mass and radius measurements. For this particular target, a more precise measurement of the effective temperature would be desirable.

Our work to provide high-precision dynamical measurements of the known \textit{Kepler} sample of eclipsing binaries with an oscillating giant from the catalogue of \citet{Gaulme2019} continues. Remaining issues for pushing the dynamical radius precision and accuracy below 1\% for these eclipsing binaries with an oscillating giant include the potential uncertainty on third light contamination and limb darkening. The problem with uncertain third light due to nearby stars that are blended with the binary due to large pixels could be addressed by potential future space data from missions such as STEP\footnote{https://space.au.dk/the-space-research-hub/step/} and/or HAYDN \citep{Miglio2021B}, which are not troubled by contamination due to large pixels. HAYDN also aims to provide both dynamical and asteroseismic masses and radii of stars in clusters, thus increasing the parameter space where comparisons can be made. Efforts are ongoing to find more targets with the TESS mission \citep[Sect. 7.1]{Prsa2022}, which could be followed up by the same missions in order to avoid contamination. However, potential intrinsic third light remains an issue that will continue to need attention.

\begin{acknowledgements}
\\      
We thank the anonymous referee for useful comments that helped improve the paper.
\\
We thank Saul A. Rappaport for help with details on the method to estimate the potential systematic error in the Gaia parallax due to the binary orbit.
\\
Based on observations made with the Nordic Optical Telescope, owned in collaboration by the University of Turku and Aarhus University, and operated jointly by Aarhus University, the University of Turku and the University of Oslo, representing Denmark, Finland and Norway, the University of Iceland and Stockholm University at the Observatorio del Roque de los Muchachos, La Palma, Spain, of the Instituto de Astrofisica de Canarias.
\\
Funding for the Stellar Astrophysics Centre is provided by The Danish National Research Foundation (Grant DNRF106). 
\\
AM acknowledges support from the ERC Consolidator Grant funding scheme (project ASTEROCHRONOMETRY, https://www.asterochronometry.eu, G.A. n. 772293).
\\
D.H. acknowledges support from the Alfred P. Sloan Foundation and the National Aeronautics and Space Administration (80NSSC19K0597).
\\

We thank D. A. VandenBerg for providing Victoria isochrones assuming specific element compositions and ages at our request.

This work has made use of data from the European Space Agency (ESA) mission
{\it Gaia} (\url{https://www.cosmos.esa.int/gaia}), processed by the {\it Gaia}
Data Processing and Analysis Consortium (DPAC,
\url{https://www.cosmos.esa.int/web/gaia/dpac/consortium}). Funding for the DPAC
has been provided by national institutions, in particular the institutions
participating in the {\it Gaia} Multilateral Agreement.
\\
This paper includes data collected by the Kepler mission. Funding for the Kepler mission is provided by the NASA Science Mission directorate.
\\
Some of the data presented in this paper were obtained from the Mikulski Archive for Space Telescopes (MAST). STScI is operated by the Association of Universities for Research in Astronomy, Inc., under NASA contract NAS5-26555. Support for MAST for non-HST data is provided by the NASA Office of Space Science via grant NNX09AF08G and by other grants and contracts.

\end{acknowledgements}


   \bibliographystyle{aa} 
   \bibliography{brogaard} 
%

\begin{appendix}

\section{Radial-velocity measurements}


\begin{table}[h]
\caption{Radial-velocity measurements of KIC\,4054905.}             
\label{table:A1}      
\begin{tabular}{l c c}        
\hline\hline                 
BJD\_TDB & $\rm RV_p (km\cdot s^{-1})$ & $\rm RV_s (km\cdot s^{-1})$ \\    
\hline                        
2459310.69531 &  38.58(23) &  -15.79(83) \\
2459314.69531 &  39.61(27) &  -16.67(57)\\
2459323.66016 &  39.11(14) &  -16.14(13)\\
2459332.63672 &  34.20(20) &  -11.64(13)\\
2459338.64063 &  28.67(20) &  -5.84(27)\\
2459380.59766 &   0.16(17) &  22.56(08)\\
2459441.44531 &  -2.90(15) &  25.61(27)\\
2459452.46484 &  -1.90(17) &  24.90(34)\\
2459464.48438 &  -0.40(07) &  23.15(49) \\
2459465.51953 &  -0.31(09) &  22.88(26)\\
2459473.52343 &   0.95(06) &  21.69(40)\\
\hline                                   
\end{tabular}
\end{table}

\end{appendix}

\end{document}